\documentclass[
onecolumn,
]{aastex631}

\usepackage{hyperref}
\usepackage{amsmath}
\usepackage{bm}

\accepted{July 28, 2024}

\submitjournal{ApJ}

\begin{document}

\title{Observations of Kappa Distributions in Solar Energetic Protons and Derived Thermodynamic Properties}

\shortauthors{Cuesta et al.}

\correspondingauthor{M. E. Cuesta}
\email{mecuesta@princeton.edu}

\author[0000-0002-7341-2992]{M. E. Cuesta}
\affiliation{Department of Astrophysical Sciences, Princeton University, Princeton, NJ 08544, USA}

\author{A. T. Cummings}
\affiliation{Department of Astrophysical Sciences, Princeton University, Princeton, NJ 08544, USA}

\author[0000-0002-7655-6019]{G. Livadiotis}
\affiliation{Department of Astrophysical Sciences, Princeton University, Princeton, NJ 08544, USA}

\author[0000-0001-6160-1158]{D. J. McComas}
\affiliation{Department of Astrophysical Sciences, Princeton University, Princeton, NJ 08544, USA}

\author[0000-0002-0978-8127]{C. M. S. Cohen}
\affiliation{Space Research Lab, California Institute of Technology, Pasadena, CA 91125, USA}

\author[0000-0003-0412-1064]{L. Y. Khoo}
\affiliation{Department of Astrophysical Sciences, Princeton University, Princeton, NJ 08544, USA}

\author[0000-0002-8527-1509]{T. Sharma}
\affiliation{Department of Astrophysical Sciences, Princeton University, Princeton, NJ 08544, USA}

\author[0000-0002-3093-458X]{M. M. Shen}
\affiliation{Department of Astrophysical Sciences, Princeton University, Princeton, NJ 08544, USA}

\author[0000-0002-6962-0959]{R. Bandyopadhyay}
\affiliation{Department of Astrophysical Sciences, Princeton University, Princeton, NJ 08544, USA}

\author[0000-0002-8111-1444]{J. S. Rankin}
\affiliation{Department of Astrophysical Sciences, Princeton University, Princeton, NJ 08544, USA}

\author[0000-0003-2685-9801]{J. R. Szalay}
\affiliation{Department of Astrophysical Sciences, Princeton University, Princeton, NJ 08544, USA}

\author[0000-0001-7952-8032]{H. A. Farooki}
\affiliation{Department of Astrophysical Sciences, Princeton University, Princeton, NJ 08544, USA}

\author[0000-0002-9246-996X]{Z. Xu}
\affiliation{Space Research Lab, California Institute of Technology, Pasadena, CA 91125, USA}

\author[0000-0003-0581-1278]{G. D. Muro}
\affiliation{Space Research Lab, California Institute of Technology, Pasadena, CA 91125, USA}

\author[0000-0002-7728-0085]{M. L. Stevens}
\affiliation{Smithsonian Astrophysical Observatory, Cambridge, MA 02138, USA}

\author[0000-0002-1989-3596]{S. D. Bale}
\affiliation{Physics Department, University of California, Berkeley, CA, 94720, USA}
\affiliation{Space Sciences Laboratory, University of California, Berkeley, CA 94720, USA}



\begin{abstract}

In this paper we model the high-energy tail of observed solar energetic proton energy distributions with a kappa distribution function. 
We employ a technique for deriving the thermodynamic parameters of solar energetic proton populations measured by the Parker Solar Probe (PSP) Integrated Science Investigation of the Sun (IS\(\odot\)IS) EPI-Hi high energy telescope (HET), over energies from 10 - 60 MeV.
With this technique we explore, for the first time, the characteristic thermodynamic properties of the solar energetic protons associated with an interplanetary coronal mass ejection (ICME) and its driven shock. 
We find that 
(1) the spectral index, or equivalently, the thermodynamic parameter kappa of solar energetic protons (\(\kappa_{\rm EP}\)) gradually increases starting from the pre-ICME region (upstream of the CME-driven shock), reaching a maximum in the CME ejecta (\(\kappa_{\rm EP} \approx 3.5\)), followed by a gradual decrease throughout the trailing portion of the CME; 
(2) solar energetic proton temperature and density (\(T_{\rm EP}\) and \(n_{\rm EP}\)) appear anti-correlated, a behavior consistent to sub-isothermal polytropic processes; and 
(3) values of \(T_{\rm EP}\) and \(\kappa_{\rm EP}\) appear are positively correlated, indicating an increasing entropy with time.
Therefore, these proton populations are characterized by a complex and evolving thermodynamic behavior, consisting of multiple sub-isothermal polytropic processes, and a large-scale trend of increasing temperature, kappa, and entropy. 
This study and its companion study by \citet{LivadiotisEA2024_SEPthermo_technique} open a new set of procedures for investigating the thermodynamic behavior of energetic particles and their shared thermal properties.

\end{abstract}



\section{Introduction} \label{sec:intro}

Solar wind plasma fills interplanetary space, extending from the solar corona to the termination shock of the heliosphere. 
As in the vast majority of space plasmas, the solar wind is characterized by long-range interactions, which induce correlations among particles over multiple temporal and spatial scales. 
These include the length of Debye shielding \citep{LivadiotisMcComas2013SSRv_KappaDist_SpSc} and length of magnetic correlation of \(\sim 10^6\)~km near 1~au \citep{FiskSara1973JGR_Correlation_1au,Hedgecock1975SoPh_Correlation_CRs,matthaeus1982JGR}. 
Due to the presence of long-range particle correlations, the frameworks of \citet{Boltzmann1866}-\citet{Gibbs1902elementary_book} (BG) entropy and the Maxwellian canonical velocity distribution \citep{Maxwell1860} fail to describe particle velocity distributions observed in space plasmas. 
Instead, the more generalized framework of kappa distributions and their associated statistical mechanics and thermodynamics is required \citep{LivadiotisMcComas2013SSRv_KappaDist_SpSc,LivadiotisMcComas2023ApJ_PolytropicIndex_Heating,livadiotis2017kappa_book}, which reduces to the BG and Maxwell representation in the limit of negligible long-range correlations \citep{LivadiotisMcComas2023ApJ_PolytropicIndex_Heating}. The statistical framework of kappa distributions conforms to 
(1) the canonical ensemble entropy maximization \citep{Tsallis1988JSP_BGgeneralize, Treumann1997GeoRL_Magnetopause_diff, MilovanovZelenyi2000NPGeo_TsallisEntropy_function, Leubner2002ApSS_NonExtKappaEntropy, LivadiotisMcComas2009JGRA_ExploitTsallisSpacePlasma,Livadiotis2014Entrp_LagrangianTemperatureKappa,JizbaKorbel2019PhRvL_MaxEntropy}, 
(2) the principles of thermodynamics \citep{Abe2001PhRvE_ComposableEntropy,Livadiotis2018Entropy,LivadiotisMcComas2021Entrp_ThermoTempKappa_EntropyDefect,LivadiotisMcComas2022ApJ_Correlations_Kappa,LivadiotisMcComas2023ApJ_PolytropicIndex_Heating}, and 
(3) a polytropic behavior \citep{LivadiotisMcComas2009JGRA_ExploitTsallisSpacePlasma}.

This generalized statistical framework has been applied to a variety of systems in nature having long-range correlations among their particles, such as space plasmas. 
In particular, kappa distributions have been successfully used to accurately describe velocity distributions in space plasmas throughout the heliosphere \citep[e.g.,][]{LivadiotisMcComas2013SSRv_KappaDist_SpSc,livadiotis2017kappa_book}. 
Solar wind plasma parameters have frequently been determined through kappa distribution modeling \citep[e.g.,][]{CollierEA1996GeoRL_HeavyDensities_SW,MaksimovicEA1997AAP_KappaCoronaSW,MaksimovicEA1997GeoRL_ElectronKappa_Ulysses,PierrardEA1999JGR_ElectronVDFs,marsch2006kinetic,Zouganelis2008JGRA_SuprathermalParams_QTNkappa,StverakEA2009JGRA_RadEvol_ElectPops,LivadiotisMcComas2010ApJ_SpacePlasmaNonEquil,Yoon2012PhPl_ElectronKappa_LangmuirTurb,Yoon2014JGRA_ElectronKappa_QTN}. 
Kappa distributions have also been used to model solar energetic particle (SEP) distributions at higher energies \citep[\(\sim\)1~keV and \(\sim\)1~MeV; e.g.,][]{Shizgal2022APSS_kappaSEPsModel}.

SEPs permeate the solar wind and can variously be accelerated by varying mechanisms; for example, SEPs can be accelerated by interplanetary shocks via turbulent diffusion acceleration processes \citep{Fisk1971JGR_CosRayIntensity_IPshock,AxfordEA1977ICRC_CosRayAccel_Shock,BlandfordOstriker1978ApJL_ParticleAccel_Shock,Bell1978MNRAS_ShockAccel_CosRay,GoslingEA1979AIPC_IonAccel_EarthBowShock,Lee1983JGR_IonAccel_IPshock,Lee1997GMS_CMEshock_transport}, seeded by the suprathermal (ST) particle population \citep[energies generally between \(\sim\)~10~keV to \(\sim\)~1~MeV per nucleon;][]{MasonGloeckler2012SSRv_suprathermals}).
Particle acceleration mechanisms can contribute to an increase of particles with energies typical of the ST tail of the distribution, thereby raising the tail of the kappa distribution. This corresponds to a decrease in the value of thermodynamic kappa \citep[e.g., see Figure 6b in][]{LivadiotisMcComas2011ApJ_InvariantKappa}.

To our knowledge, the thermodynamics of SEPs have remained uninvestigated. 
\citet{LivadiotisMcComas2011ApJ_InvariantKappa,LivadiotisMcComas2012ApJ_KappaThermoProcesses,LivadiotisMcComas2013SSRv_KappaDist_SpSc,LivadiotisMcComas2023ApJ_PolytropicIndex_Heating} provide the detailed derivation of a technique for the application of this generalized thermodynamic framework of kappa distributions, which was adapted by \citet{LivadiotisEA2024_SEPthermo_technique} for SEPs. 
Here we briefly summarize this technique and apply it to solar energetic protons over energies of \(\sim\)10 to 60 MeV measured on the Parker Solar Probe (PSP) \citep{fox2016SSR} by the Integrated Science Investigation of the Sun (IS\(\odot\)IS) instrument suite \citep{mccomas2016SSR}. 
In this paper, we use this procedure to examine the temporal profile of the thermodynamic quantities for an SEP event associated with an interplanetary coronal mass ejection (ICME) observed by PSP, for the first time. 
We track the changes in thermal pressure associated with the energetic proton population via variations in their density and temperature, which can also be used to explore their polytropic and entropic behavior in the different regions of an ICME.

The paper is organized as follows. 
In Section \ref{sec:data_method}, we describe the observational data of solar energetic protons and the procedure used to compute the thermodynamic parameters of these protons.  
In Section \ref{sec:results}, we derive the spectral index (which coincides with the thermodynamic parameter \(\kappa\)), the temperature (\(T\)), and the density (\(n\)), from fitting kappa distributions to the observed spectra of solar energetic protons. 
In Section \ref{sec:discussion}, we compute the polytropic index and entropic measure, characterizing the thermodynamics of the involved polytropic processes and their evolution, and finally, discuss the implications of our results from a thermodynamic perspective. 
In Section \ref{sec:conclusion}, we summarize the findings of this study. 
In the Appendix \ref{app:entropy}, we provide details concerning the relationship between an entropic measure and thermodynamic kappa.

\section{Data and Methods} \label{sec:data_method}

\citet{LivadiotisEA2024_SEPthermo_technique} derive the detailed procedures for extracting thermodynamic quantities, such as the parameter \(\kappa\), density, and temperature, from the intensity spectra of energetic protons.
Here, we briefly summarize the core elements of this procedure, starting from the 3D kappa distribution of the proton velocities given by \citep{LivadiotisMcComas2013SSRv_KappaDist_SpSc,livadiotis2017kappa_book}:

\begin{equation}\label{eq:3d_velocity_distribution}
    P_{u}(\bm{u})=\pi^{-\frac{3}{2}}\cdot {\rm N}(\kappa_0)\cdot \theta^{-3}\cdot \left[1+ \frac{1}{k_0}\frac{(\bm{u}-\bm{u}_b)^2}{\theta^2}\right]^{-\kappa_0 - \frac{5}{2}},
\end{equation}
which involves the particle \(\bm{u}\) and bulk \(\bm{u}_b\) proton velocities.
The temperature of the protons is expressed in terms of the speed-scale parameter \(\theta=\sqrt{2T/m}\), where \(m\) is the proton mass and \(T\) is the proton temperature in energy units.
The factor \({\rm N}(\kappa_0)\equiv\kappa_0^{-\frac{3}{2}}\cdot\frac{\Gamma(\kappa_0+\frac{5}{2})}{\Gamma(\kappa_0+1)}\) is a characteristic function of the invariant thermodynamic kappa parameter, \(\kappa_0=\kappa -\frac{3}{2}\), with \({\rm N}(\kappa_0\rightarrow\infty)=1\), representing the factor in the limit of classical thermal equilibrium.
By expressing Eq. \eqref{eq:3d_velocity_distribution} in terms of the kinetic energy as measured in the comoving frame, \(\epsilon=\frac{1}{2}m\left(\bm{u}-\bm{u}_b\right)^2\), taking its logarithm (base 10), and expanding up to the parabolic term, we obtain:

\begin{align}\label{eq:logP_expanded}
    \begin{split}
        \log P_u(\epsilon) \cong \log\left[ \pi^{-\frac{3}{2}} \cdot {\rm N}(\kappa_0) \cdot \left(\frac{2T}{m}\right)^{-\frac{3}{2}}\right]+\left(\kappa_0+\frac{5}{2}\right) \cdot \log \left[\frac{\kappa_0 T}{\left[{\rm keV}\right]}\right] \\ 
        - \left(\kappa_0+\frac{5}{2}\right) \cdot \log\left[\frac{\epsilon}{[{\rm keV}]}\right]-\frac{(\log e)\cdot\kappa_0\left(\kappa_0+\frac{5}{2}\right)\cdot T}{\epsilon},
    \end{split}
\end{align}
where the last term of Eq. \eqref{eq:logP_expanded} can be ignored in the limit \(\epsilon \gg \kappa_0 T\), reducing the kappa distribution into a power-law relationship.
This enables us to model the tail of the observed flux-energy spectrum as a power-law with fitting parameters related to the terms in Eq. \eqref{eq:logP_expanded}.
The reduced power-law has two fitting parameters (defined below) that involve the thermodynamic parameters of density, temperature, and kappa.
The functional dependence of these two power-law tail fitting parameters on the thermodynamic parameters comprises a special expression, characteristic to the formulation of the kappa distribution.
Next, we derive the expression that is used for examining whether the tail indeed originated from a kappa distribution function.
If yes, we can proceed to determine the thermodynamic parameters involved in the original kappa distribution.
This technique was previously developed by \citet{LivadiotisMcComas2011ApJ_InvariantKappa,LivadiotisMcComas2012ApJ_KappaThermoProcesses,LivadiotisMcComas2013SSRv_KappaDist_SpSc,LivadiotisMcComas2023ApJ_PolytropicIndex_Heating} and later adapted by \citet{LivadiotisEA2024_SEPthermo_technique} for SEP flux-energy spectra.

Following this technique, we fit the spectrum of the energetic proton intensity with energy modeled as the tail of an assumed kappa distribution (Eq. \eqref{eq:logP_expanded}).
Consequently, the thermodynamic parameters become \(\kappa \rightarrow \kappa_{\rm EP}\), \(n \rightarrow n_{\rm EP}\), and \(T \rightarrow T_{\rm EP}\), where the subscript `EP' corresponds to an energetic particle population (in this study, only the intensity of energetic protons is analyzed).
Under the limit \(\epsilon \gg \kappa_0{\rm k_B}T\), the tail of the spectrum is a power-law that can be fitted by the form:

\begin{equation}\label{eq:fitfunc1}
    {\rm Y}^{(1)} \cong {\rm logInt^{(1)}} + {\rm Slo^{(1)}} \cdot {\rm X}^{(1)},
\end{equation}
where we set the variables as 

\begin{equation}\label{eq:vars1}
    {\rm X^{(1)}} \equiv \log \left[ \frac{\epsilon}{[{\rm keV}]}\right] \text{ ; } {\rm Y^{(1)}} \equiv \log \left[ \frac{J}{[{\rm cm^{-2}\cdot s^{-1} \cdot sr^{-1} \cdot keV^{-1}}]} \right],
\end{equation}
with \(J\) as the particle intensity, for the following fitting parameters, intercept (\({\rm logInt}^{(1)}\)) and slope (\({\rm Slo}^{(1)}\)):

\begin{align}\label{eq:params1}
\begin{split}
    {\rm logInt^{(1)}} &\equiv C + \log \left[\frac{n}{[{\rm cm^{-3}}]}\right] + (\kappa_0 + 1)\cdot \log \left[\frac{T}{[{\rm keV}]}\right] + \log \left[\kappa_0^{\kappa_0 + 1}\cdot \frac{\Gamma\left(\kappa_0+\frac{5}{2}\right)}{\Gamma\left(\kappa_0 + 1\right)}\right], \\
    {\rm Slo}^{(1)} &\equiv - \left(\kappa_0 +\frac{3}{2}\right) = - \kappa.
\end{split}
\end{align}
Note that the absolute slope (or, the spectral index) equals the thermodynamic kappa, \(\kappa\).
In the above expressions, 
\(n\) is the proton density and the constant \(C\) is given by:

\begin{equation}\label{eq:constant}
    C = \log \left[2^{-\frac{1}{2}} \cdot \pi^{-\frac{3}{2}} \left(\frac{m}{[{\rm kg}]}\right)^{-\frac{1}{2}}\left(\frac{[{\rm keV}]}{[{\rm j}]}\right)^{\frac{1}{2}} \frac{[{\rm m}]}{[{\rm cm}]}\right] = \log \left[ \frac{100}{\sqrt{2\pi^3m}} \left( \frac{{\rm [keV] \cdot [kg]}}{{\rm [j]}}\right)^{\frac{1}{2}} \right] \cong 6.60-\frac{1}{2}\log \left[m_{\rm amu}\right]
\end{equation}
after accounting for the proper units and proton mass, where \(m_{\rm amu}\) is the atomic mass of the particle (\(m_{\rm amu} =1\) for protons).
When \({\rm Slo^{(1)}}\) increases/decreases, the tail of the intensity spectrum steepens/flattens, corresponding to a larger/smaller thermodynamic \(\kappa\) or, more generally, an increased/decreased entropy.

For each one minute sample, we fit the variables \({\rm X^{(1)}}\) and \({\rm Y^{(1)}}\) according to Eq. \eqref{eq:fitfunc1}, enabling us to collect a time series of values for \({\rm logInt^{(1)}}\) and \({\rm Slo^{(1)}}\), each dependent on \(\kappa_{\rm EP}\), the thermodynamic kappa for energetic particles.
Because of the formulation of kappa distributions and their inherent power-law tails, the factor \(\log \left[ {\rm A}(\kappa_{\rm EP}) \right]\) organizes a linear relationship between a new set of variables, \({\rm X^{(2)}}\) and \({\rm Y^{(2)}}\), which allows us to derive the energetic proton density (\(n_{\rm EP}\)) and temperature (\(T_{\rm EP}\)).
It is this linear relationship between \({\rm X^{(2)}}\) and \({\rm Y^{(2)}}\) that signifies the presence of kappa distributed velocities \citep{LivadiotisMcComas2011ApJ_InvariantKappa,LivadiotisMcComas2022ApJ_Correlations_Kappa}.
Following the definitions of \({\rm logInt^{(1)}}\) and \({\rm Slo^{(1)}}\) in Eq. \eqref{eq:params1}, the new set of variables can be defined as:

\begin{equation}\label{eq:vars2}
    {\rm X^{(2)}} \equiv \kappa_{\rm EP} -\frac{1}{2} \text{ ; } {\rm Y^{(2)}} \equiv {\rm logInt^{(1)}} - \log\left[ {\rm A}(\kappa_{\rm EP})\right],
\end{equation}
with 

\begin{equation}\label{eq:Afunc}
    {\rm A}(\kappa_{\rm EP}) \equiv \left(\kappa_{\rm EP} -\frac{3}{2}\right)^{\kappa_{\rm EP} -\frac{1}{2}}\cdot \frac{\Gamma(\kappa_{\rm EP}+1)}{\Gamma(\kappa_{\rm EP}-\frac{1}{2})},
\end{equation}
which is derived form a variation of the spectral index.
For the new set of variables we expect a linear behavior given by:

\begin{equation}\label{eq:fitfunc2}
    {\rm Y^{(2)}} \cong {\rm logInt^{(2)}} + {\rm Slo^{(2)}}\cdot {\rm X^{(2)}},
\end{equation}
with new fitting parameters, \({\rm logInt^{(2)}}\) and \({\rm Slo^{(2)}}\), from which we can derive \(n_{\rm EP}\) and \(T_{\rm EP}\), respectively, such that

\begin{equation}
    {\rm logInt^{(2)}} \equiv C + \log \left( \frac{n_{\rm EP}}{{\rm \left[cm^{-3}\right]}}\right) \text{ ; } {\rm Slo^{(2)}} \equiv \log \left( \frac{T_{\rm EP}}{{\rm \left[keV\right]}}\right)
\end{equation}
or

\begin{equation}
    n_{\rm EP}=2.51\times 10^{-7} \times {\rm Int^{(2)}} {\rm \left[cm^{-3}\right]} \text{ ; } T_{\rm EP}=10^{\rm Slo^{(2)}} {\rm \left[keV\right]},
\end{equation}
where \({\rm Int^{(2)}}=10^{\rm logInt^{(2)}}\).

For the results shown in Section \ref{sec:results}, we choose to slide a centered 31-minute duration interval by every 1-minute sample to compute a smoothed time series for \(T_{\rm EP}\), \(n_{\rm EP}\), and \(\kappa_{\rm EP}\).
One limitation of this analysis is the sometimes-low count rates of SEPs over the selected energy range (10 to 60~MeV).
The values of (\(T_{\rm EP}\), \(n_{\rm EP}\), and \(\kappa_{\rm EP}\) in such intervals that lack sufficient counting statistics are filled as not-a-number (NaN) and reflected by gaps in the derived time series.
We do not include any value of \(\kappa \leq 1.5\) for any time within the selected interval in the second fit (Eq. \eqref{eq:fitfunc2}) used to compute \(T_{\rm EP}\) and \(n_{\rm EP}\).
This is due to the functional behavior \({\rm A}(\kappa_{\rm EP})\) near \(\kappa_{\rm EP}=\frac{3}{2}\) where \(\kappa_{\rm EP} \in \left(\frac{3}{2},\infty\right)\), which is the limiting range of possible stationary-state plasma distributions out of equilibrium \citep{LivadiotisMcComas2009JGRA_ExploitTsallisSpacePlasma,LivadiotisMcComas2010ApJ_SpacePlasmaNonEquil}.

Another limitation to this procedure of deriving a temperature and density from the spectrum of energetic protons is determining its application to intervals with significant transitions in the distribution of the proton population, for example, sampling more than one kappa distribution or sampling a non-stationary region. 
The method requires stationarity and its applicability to intervals weakens in the presence of transitions. 
Therefore, we apply a metric to determine when the procedure may be applied and if the density and temperature derived from the method may be used to understand the thermodynamic properties of energetic protons. 
Following the second fit over the basis of \({\rm X}^{(2)}\) and \({\rm Y}^{(2)}\) values (using Eq. \eqref{eq:fitfunc2}), we take sequential pairs from the same basis and directly solve for \({\rm logInt}^{(2)}\) and \({\rm Slo}^{(2)}\) resulting in no more than 30 values per 31-minute duration interval. 
Then we compute the weighted average of these solved values, denoted as \(\langle {\rm logInt}_{sol}^{(2)}\rangle\) and \(\langle {\rm Slo}_{sol}^{(2)}\rangle\). 
The t-value of these averaged values and the fitted values will then be used as our metric of applicability of the method, such that 
\(t_p=\left|\langle p_{sol} \rangle- p_{fit} \right|/\sqrt{(\sigma_{p,sol}^2+\sigma_{p,fit}^2)}\), where \(p\) is one of either \({\rm logInt}^{(2)}\) or \({\rm Slo}^{(2)}\) and \(\sigma_p\) is its error. 
We can reliably apply the method when the t-value is less than or equal to 1.7, weakly apply the method when the t-value is between 1.7 and 2.5, and poorly apply the method when the t-value is greater than 2.5. These three categories are coded to the colors green, orange, and red below, respectively, when plotting the thermodynamic parameters over time. 
All results presented this study are produced from the subset of derived parameters only under strong applicability of the procedure above. This testing has been demonstrated rigorously by \citet{LivadiotisEA2024_SEPthermo_technique}.

Uncertainties are propagated through the fits via ordinary least squares for the first fit using Eq. \eqref{eq:fitfunc1} and the orthogonal distance regression (ODR) method for the second fit using Eq. \eqref{eq:fitfunc2} \citep{BoggsRogers1990book_ODRfitting}. 
ODR fitting accepts uncertainties in both the dependent and independent variables. 
In the first fit, the intensity uncertainty comes from the statistical uncertainty in the measured intensity for a given energy bin. 
We omit uncertainty in the energy channels of EPI-Hi, where each energy is the geometric mean of the corresponding energy bin edges.
The best-fit parameters and standard errors from the first fitting are then used as input for the second linear ODR fitting to derive a temperature and density, with associated standard errors. 
The \(R^2\) values for the second fitting are also used as a filter to mask intervals with an undesirable goodness of fit in the linearity between the variables \({\rm X}^{(2)}\) and \({\rm Y}^{(2)}\), which signifies the presence that the particle population’s velocity are kappa distributed. 
We reject any interval with a value of \(R^2<0.97\) in the second fitting for the results shown below.

For this study, we utilize energetic proton data measured by the PSP/IS\(\odot\)IS EPI-Hi HET \citep{mccomas2016SSR}, available at NASA’s Space Physics Data Facility\footnote{NASA's Space Physics Data Facility can be found at \href{https://cdaweb.gsfc.nasa.gov/}{https://cdaweb.gsfc.nasa.gov/}.}. 
We focus on data measured by HET-A, the sunward facing aperture of HET on EPI-Hi \citep{WiedenbeckEA2021AandA_PSP_ISOIS_HET} that is mounted to be centered along a line \(20\deg\) from the spacecraft-Sun line (\(45\deg\) from this line is close to a nominal Parker spiral direction for a 400~km/s solar wind speed). 
We examine protons with energies from 10 to 60~MeV for a SEP event associated with a CME-driven shock during the time range 15 Feb 2022 12:00:00 to 19 Feb 2022 00:00:00 at \(\approx\)~0.35~au. 
For more details regarding the CME structure and multi-spacecraft observations for the 15 Feb 2022 SEP event, see \citet{PalmerioEA2024ApJ_CME_BepiPSP_2022feb15} and \citet{KhooEA2024ApJ_CME_BepiPSP_2022feb15}. 
We also use 1-minute magnetic field data measured by PSP/FIELDS \citep{bale2016SSR} to contextualize the data in terms of the magnetic environment encountered by PSP. 
Finally, the SWEAP/SPAN-I \citep{kasper2016SSR,LiviEA2022ApJ_PSP_SPAN} instrument provides bulk proton temperature and density, enabling side-by-side comparison of the thermodynamic behavior between protons described by the thermal core (bulk) and those described by the distribution far from the thermal core (energetic).
All quantities derived from SPAN-I are smoothed by a 10-point rolling median.
The gyro-radius for 10 and 60 MeV protons, \(\rho_p\), are computed for reference, such that \(\rho_p=\nu_{\rm Ti}/\omega_{\rm ci}\) where \(\nu_{\rm Ti}\) is the ion trapping rate and \(\omega_{\rm ci}\) is the ion gyro-frequency.

\section{Results} \label{sec:results}

Having outlined the formalism to investigate SEP thermodynamics, we analyze the evolution of solar energetic proton thermodynamic properties for an SEP event as measured by PSP/IS\(\odot\)IS EPI-Hi HET-A. 
Figure 1 shows a sample intensity-energy spectrum at 06:24:30 on 16 Feb 2022 (left panel), and the derived \(T_{\rm EP}\) and \(n_{\rm EP}\) extracted from the fit to the compiled values of \({\rm X}^{(2)}\) and \({\rm Y}^{(2)}\) from a 31-minute interval from 06:09:30 to 06:40:30 (right panel). 
This suggests that a power-law fit to the intensity-energy spectrum is a suitable choice, and that the kappa distribution function can be used to describe the intensity spectra measured by HET-A for the observed energy range, as found in \citet{LivadiotisEA2024_SEPthermo_technique}. We also show the expected linear relationship between the variables \({\rm X}^{(2)}\) and \({\rm Y}^{(2)}\), which confirms that the velocity distribution of the observed energetic protons are well described by a kappa distribution function in the high energy limit (a power-law behavior when \(\epsilon \gg \kappa_0 T_{\rm EP}\)). 
For the discussion below, an increase/decrease in \(\kappa_{\rm EP} = \kappa_0+3/2=-{\rm Slo}^{(1)}\) corresponds to a steeper/shallower spectrum \citep{DayehLivadiotis2022ApJL_PolytropicBulk_ICMEs}. 

\newpage

\begin{figure}[!ht]
    \centering
    \includegraphics[width=.8\textwidth]{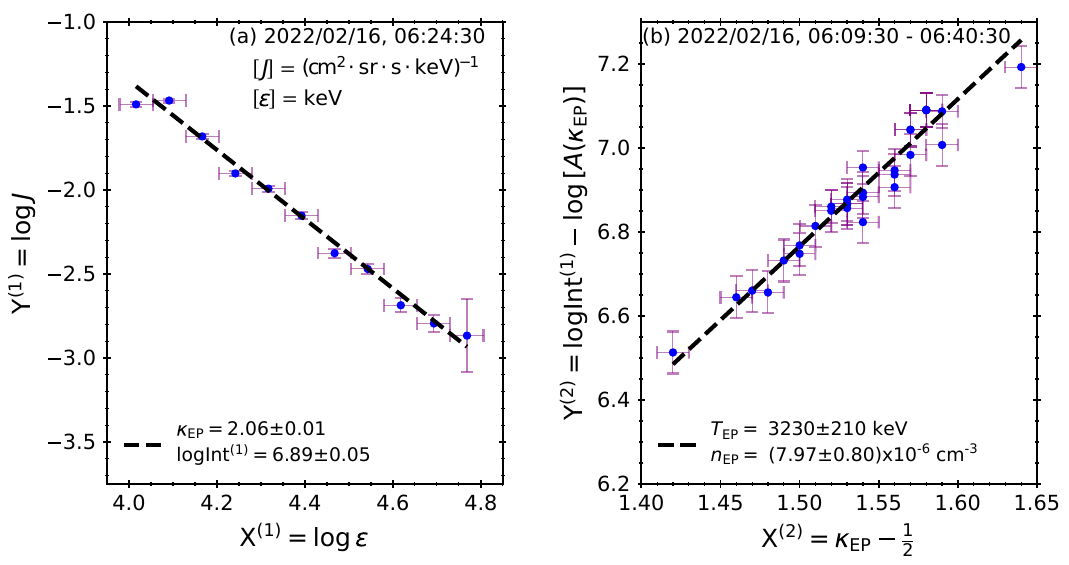}
    \caption{A sample energy spectrum measured with EPI-Hi HET-A on 16 Feb 2022 06:24:30 with its corresponding power-law fit (left) and the compiled fitted values of the left panel over the interval 16 Feb 2022 06:09:30 to 06:40:30 (right). The units of flux \(J\) and energy \(\epsilon\) are \({\rm cm^{-2} \cdot sr^{-1} \cdot s^{-1} \cdot keV^{-1}}\) and keV, respectively.  In the left panel, vertical error bars correspond to the statistical uncertainty in the measured flux at that given energy bin and horizontal error bars correspond to the energy bin widths. In the right panel, vertical and horizontal error bars represent the standard errors propagated from the ODR power-law fit to Eq. \eqref{eq:fitfunc1}. The resulting derived temperature and density are given in the legend on the right. All error bars are represented in logarithmic space.
    }
    \label{fig:sample_spectrum_fit}
\end{figure}

We then apply the same procedure to observations of 10 to 60 MeV protons to examine the evolution of their thermodynamic properties throughout the 15 Feb 2022, SEP event. 
The time series of the derived EP temperature, density, thermal pressure (\(n_{\rm EP} \cdot T_{\rm EP}\)), and parameter \(\kappa_{\rm EP}=\kappa_0+3/2\) (or spectral index) are given in Figure 2, in addition to an intensity spectrogram of the 10 – 60 MeV protons, magnetic field, SW temperature, density, and pressure for reference. In the pre-ICME region, \(\kappa_{\rm EP}\) gradually increases (the spectrum steepens) throughout the period of velocity dispersion leading to the CME-driven shock (vertical dashed red line in Figure 2). 
The CME-driven shock and CME ejecta arrival times for the 15 Feb 2022 SEP event are given by \citet{PalmerioEA2024ApJ_CME_BepiPSP_2022feb15} and \citet{KhooEA2024ApJ_CME_BepiPSP_2022feb15}. The derived temperature and density from the observed proton population are most variable in the pre-ICME, sheath, and CME ejecta regions. After the CME is encountered, the derived energetic particle temperature and spectral index are nearly constant, whereas the density continues to gradually decrease. We note that the applicability of the method is more consistently weak in the sheath region, which may be expected due its strong turbulent state resulting in a mixture of states, or more transitions, observed by PSP.

\newpage

\begin{figure}[!ht]
    \centering
    \includegraphics[width=\textwidth]{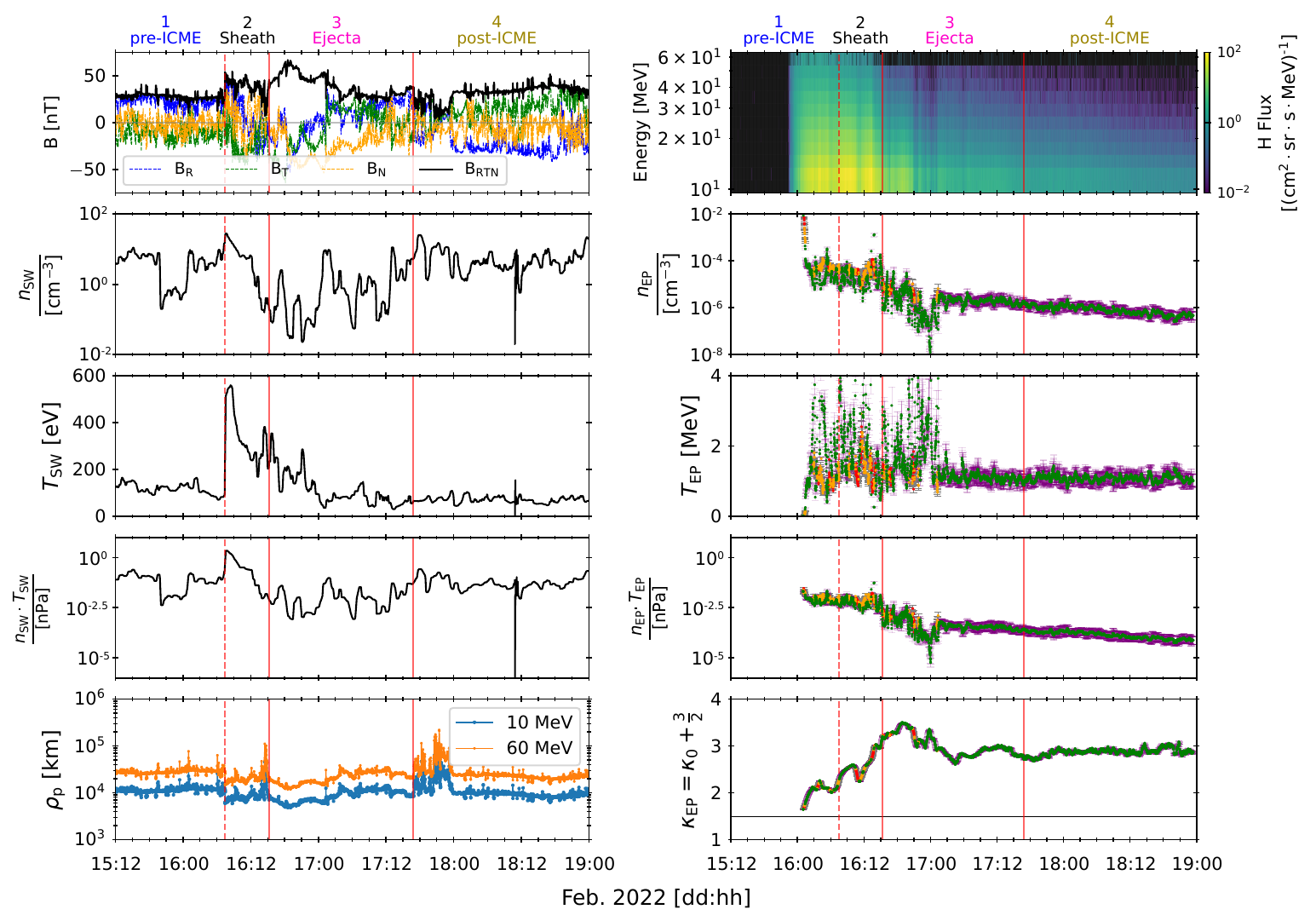}
    \caption{Overview of plasma and SEP conditions for the 15 Feb 2022 SEP event. The left panels are magnetic field RTN-components and magnitude (blue, green, orange, and black, respectively), SW proton density, SW proton temperature, SW proton thermal pressure, and gyroradii for 10 and 60 MeV protons (blue and orange) in descending order. The right panels are HET-A intensity spectrogram, EP density, EP temperature, EP thermal pressure, and EP thermodynamic kappa in descending order. Error bars represent the propagated standard errors through the ODR fitting method (see Section \ref{sec:data_method}), and are represented in logarithmic space. Colors of the derived EP variables are coded to the applicability of the method, such that green is good, orange is okay, and red is bad. In all panels, the vertical red dashed line indicates the CME-driven shock arrival time and the region between the vertical red solid lines indicates the CME ejecta. Each portion of the SEP event is labeled along the top axis as pre-ICME (solar wind upstream of the shock), sheath (downstream of the shock), CME ejecta (magnetic obstacle), and post-ICME (solar wind after the ejecta).}
    \label{fig:overview}
\end{figure}

\begin{figure}[!ht]
    \centering
    \includegraphics[width=\textwidth]{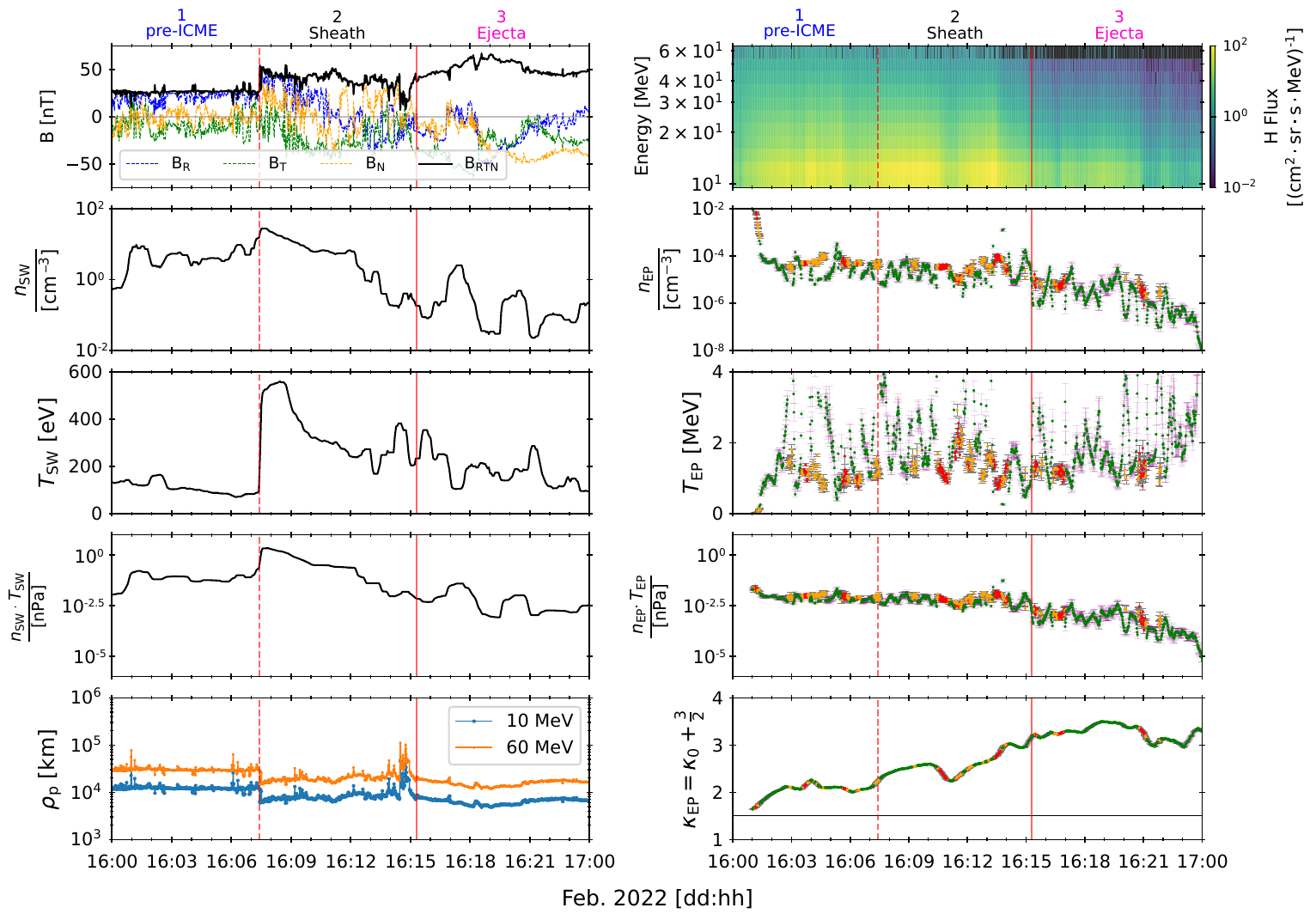}
    \caption{Same as Figure \ref{fig:overview}, zoomed into the regions near the CME-driven shock and CME ejecta arrival times.}
    \label{fig:zoomed}
\end{figure}

An interesting feature is the anti-correlation between the EP temperature and density. 
To view this behavior more clearly, Figure \ref{fig:zoomed} zooms into a region of the most variable temperature and density from Figure \ref{fig:overview}. 
During these large variations, a clear rise/fall in \(T_{\rm EP}\) corresponds to a fall/rise in the \(n_{\rm EP}\). 
We also note that \(\kappa_{\rm EP}\) is smoothly varying throughout this period of highly variable temperature and density, possibly indicating a sharing of thermodynamic information between these complex regions of otherwise separate plasma parcels.

\newpage

\section{Discussion} \label{sec:discussion}

In Figure \ref{fig:overview}, \(\kappa_{\rm EP}\) starts from a low value that gradually increases with minor deviations until PSP encounters the CME-driven shock (vertical dashed red line in Figure \ref{fig:overview}), during which \(\kappa_{\rm EP}\) increases consistently overall with a significant reduction in the middle of the sheath region. 
The spectrum continues to steepen into the CME ejecta given by the first vertical solid red line in Figure \ref{fig:overview}, the region where \(\kappa_{\rm EP}\) achieves its peak value (\(\kappa_{\rm EP} \approx 3.5\)).
The spectrum then gradually decreases throughout the latter portions of the CME. 
This suggests that the CME ejecta is the region of the highest entropy-state compared to other portions of the SEP event.  
Also, \(\kappa_{\rm EP}\) is smoothly varying during times of highly variable \(T_{\rm EP}\) and \(n_{\rm EP}\). 
This might be unexpected since a smoothly varying parameter \(\kappa_{\rm EP}\) implies that the thermodynamic behavior of these particles is being shared across all regions of the observed SEP event, i.e, the trend in \(\kappa_{\rm EP}\) suggests an ordering in the thermodynamic behavior of energetic protons.

How is it that the thermodynamic behavior exhibits a consistent evolution across the complex structure of an ICME? 
One possibility is that the energetic particles can mix across the structure such that their thermodynamic properties are shared. 
To investigate this further, we examine the gyroradius of protons (\(\rho_p\)) with energies from 10 to 60 MeV, given in Figures 2 and 3. We estimate \(\rho_p \sim\)~30,000~km at \(\approx\)~0.35~au during this SEP event, which is on the order of the scale size of supergranules on the solar surface \citep{Hart1956MNRAS_GranuleScales} where the 60~MeV \(\rho_p\) would be much smaller than the one estimated at 0.35~au. 
This means that at no point during the transit of these energetic protons did they ever significantly traverse beyond their immediately bound magnetic environment due to the radial expansion of field lines.  
This gives us reason to suspect other processes are the cause, such as the energetic particles’ transport involving mean free path lengths or perpendicular diffusion. 
The mean free path length for protons of 10 MeV or greater is estimated to range from 0.016 to 0.149 au \citep{LiEA2021ApJ_MeanFreePaths_EPs}, which may be large enough for an adequate mixing of particles with varying thermodynamic properties.

Alternatively, even if the particles are trapped within elements of the complex structure, the turbulent nature of the structure may allow the mixing of particles.
\citet{PecoraEA2021MNRAS_HelicalEPboundary} showed that small-scale magnetic flux ropes (SMFRs) can act as boundaries for energetic particles. 
\citet{FarookiEA2024ApJS_SMFRs_at1au} found that a significant portion of the solar wind is constantly filled with SMFRs, even within ICME plasma. 
It follows that SMFRs are likely to play a central role in solar wind turbulence. 
By evaluating the heliospheric evolution of the axial magnetic flux contained within SMFRs, \citet{FarookiEA2024ApJL_AxialFluxEvolution} found that axial flux increases significantly with heliocentric distance. This was interpreted as evidence that small SMFRs are statistically likely to merge forming larger flux ropes, a phenomenon previously observed in a case study by \citet{HuEA2019JPhCS_SMFRsEvolution}. 
The turbulent merging of flux ropes is a possible candidate as the mechanism responsible for the mixing of energetic particles; when two flux ropes merge their field lines, their separately trapped particles are allowed to mix. If this process occurs frequently enough, then it may result in the smoothly varying \(\kappa_{\rm EP}\) observed in this study with fluctuations that can be explained by the variation of the energetic particle content within the individual flux ropes. 

An interpretation of the overall trends in \(\kappa_{\rm EP}\) might be related to an observer’s distance to the source of acceleration processes responsible for the observed distribution of SEPs. 
In Figures \ref{fig:overview} and \ref{fig:zoomed}, \(\kappa_{\rm EP}\) is increasing closer to the CME-driven shock, which can be a source for diffusive shock acceleration \citep[][and references therein]{Lee1997GMS_CMEshock_transport}, with a noticeable brightening in the spectrograms after the shock is encountered. Once the shock is encountered, \(\kappa_{\rm EP}\) is nearly constant (gradually increases) close to the shock in the downstream turbulent sheath. 
This may similarly be explained by the modulation and/or acceleration of particles downstream of the shock due to the emergence of coherent structures and turbulence processes \citep[see][]{tessein2013ApJ,tessein2015apj,KhabarovaEA2015ApJ_MagIslands_acceleration,KhabarovaZank2017ApJ_EPs_CurrentSheets,MalandrakiEA2019ApJ_EPs_CurrentSheets,BandyopadhyayEA2020ApJS_PVI_intensity,CuestaEA2024ApJ_CMEshockTempIntensityCorrelation}. 
These processes can occur throughout the downstream plasma, resulting in the period of nearly constant spectral indices in the first half of the sheath region.
However, variations of \(\kappa_{\rm EP}\) in the pre-ICME and sheath regions may also be related to changes in the shock’s parameters during its transit. Also, the above reasoning does not explain the second larger increase of \(\kappa_{\rm EP}\) leading into the CME ejecta (the region between the vertical solid red lines in Figure \ref{fig:overview}).  
Further investigation into these observations and reasoning is required to resolve the coupling between the thermodynamic \(\kappa_{\rm EP}\) and other plasma and thermodynamic properties.

The evolution of the spectral index has been examined in many events, such that one would expect \(\kappa_{\rm EP}\) to increase as an observer approaches a shock. This can be explained by the general diffusion of the highest energy particles thus lowering their intensities, while the lowest energy particles are still locally accelerated by the shock thus raising their intensities. At and after the shock, one would expect a spectral index consistent with the shock parameters \citep{Lee1983JGR_IonAccel_IPshock} and then a period of decreasing overall intensities uniformly with energy (i.e, an expanding reservoir filled with invariant spectral shapes and decreasing intensities over time \citep{Reames2013SSRv_SEPsources}). Without such a reservoir, one might expect the spectral index to continue increasing as more high energy particles are lost. However, the CME ejecta may exhibit a different spectral shape and evolution based on its magnetic structure \citep[e.g.,][]{LarioEA2021ApJ_SEPspectra_Nov29_2020}.

\begin{figure}[!ht]
    \centering
    \includegraphics[width=.6\textwidth]{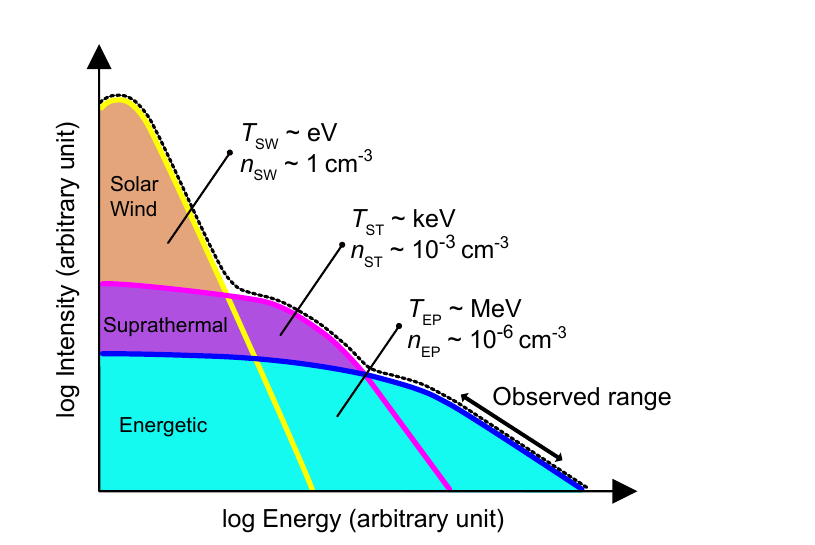}
    \caption{An illustration depicting the order of magnitudes of densities and temperatures for particle populations, and their distributions, pertaining to the solar wind (\(T_{\rm SW} \sim {\rm eV}\); \(n_{\rm SW} \sim 1~{\rm cm^{-3}}\)), suprathermal particles (\(T_{\rm ST} \sim {\rm keV}\); \(n_{\rm ST} \sim 10^{-3}~{\rm cm^{-3}}\)), and SEPs (\(T_{\rm EP} \sim {\rm MeV}\); \(n_{\rm EP} \sim 10^{-6}~{\rm cm^{-3}}\)).}
    \label{fig:illustration}
\end{figure}

We understand that the values of \(n_{\rm EP}\) are low, which are on the order of what might be expected for a constant pressure argument between populations with different energies, as schematically portrayed in Figure \ref{fig:illustration}. 
With each decreasing order of magnitude in the particle population’s density, its temperature increases an order of magnitude maintaining pressure balance between the different particle populations.  Therefore, the values of \(n_{\rm EP}\) in the present analysis fit reasonably well for the values of \(T_{\rm EP}\) derived from the 10 to 60~MeV energetic proton population observed by PSP.  
Furthermore, our analysis assumes that we are measuring a significant density of the energetic population portrayed by the blue distribution in Figure \ref{fig:illustration}. 
For this we compute the partial density of particles for the example distribution in the left panel of Figure \ref{fig:sample_spectrum_fit} between minimum energy \(\epsilon_i\) to maximum energy \(\epsilon_f\) as \(\sim \int_{\epsilon_i}^{\epsilon_f} J(\epsilon)\cdot \epsilon^{-1/2} d\epsilon\), such that \(J(\epsilon)\sim \epsilon^{1/2}p(\epsilon)g(\epsilon)\) where \(p(\epsilon\) is the phase space distribution and \(g(\epsilon)\sim \epsilon^{1/2}\) is the three-dimensional density of energy states \citep{LivadiotisMcComas2009JGRA_ExploitTsallisSpacePlasma}.
Therefore, the fractional density of particles less than 5 MeV is \(\approx 0.177\) meaning that 17.7\% of energetic protons lie below 5 MeV within the distribution, or \(\approx 82.3\%\) of energetic protons have energies greater than 5 MeV.  
This supports the utilization of the observed energy range of protons to describe the bulk of the energetic proton distribution.

\newpage

The variations in the EP temperature and density are largely anti-correlated. 
Physically, this property is the result of sub-isothermal polytropic processes. 
This behavior is investigated below. Furthermore, the energetic proton thermal pressure (\(P_{\rm EP} = n_{\rm EP} \cdot T_{\rm EP}\)), as shown in Figures \ref{fig:overview} and \ref{fig:zoomed}, tends to be nearly constant in the beginning of the SEP event leading to the arrival of the CME ejecta. 
Local pressure variations appear to respond in favor of variations in the density since the changes in density are proportionately larger than variations in temperature. 
The overall trend of the derived thermal pressure gradually decreases throughout the SEP event. 
Additionally, values of the derived thermal pressure of energetic protons are consistent with values provided by \citet{LarioEA2021ApJ_SEPspectra_Nov29_2020}.

\begin{figure}[!ht]
    \centering
    \includegraphics[width=.6\textwidth]{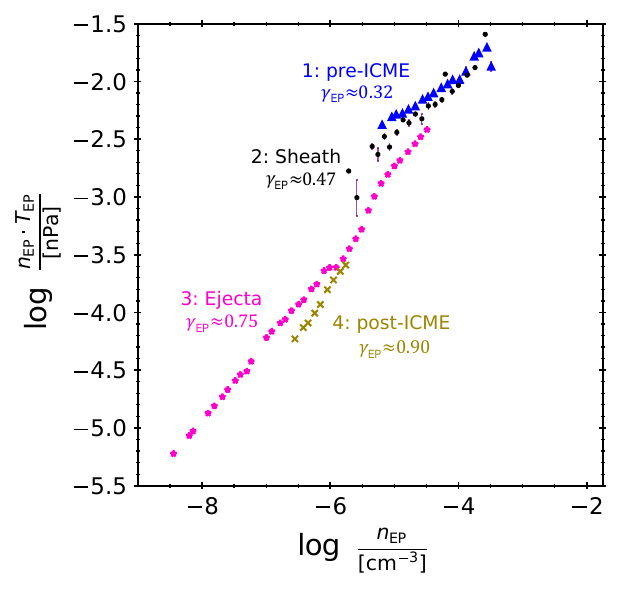}
    \caption{Diagram of \(\log P_{\rm EP}\) as a function of \(\log n_{\rm EP}\), whose variables are collapsed over time  for the portions marked along the top axis of Figure \ref{fig:overview} for the 15 Feb 2022 SEP event. These points represent the weighted average of equidistant bins of size 0.1 along \(\log n_{\rm EP}\) with standard error bars given in purple.}
    \label{fig:polytrope}
\end{figure}

The results presented here enable us to view the temporal evolution of EP thermodynamic properties, for the first time, such as their polytropic index \(\gamma_{\rm EP}\) and entropic measure \(S_{\rm EP} \equiv \ln T_{\rm EP}-\frac{2}{3}\ln n_{\rm EP}\).
A useful relation to compute \(\gamma_{\rm EP}\) is the thermodynamic relation between thermal pressure and density for polytropes, given as

\begin{equation} \label{eq:polytrope}
    \log \left[ \frac{P_{\rm EP}}{{\rm \left[Pa\right]}} \right] \sim \gamma_{\rm EP} \cdot \log \left[ \frac{n_{\rm EP}}{{\rm \left[cm^{-3}\right]}} \right] + {\rm const.}
\end{equation}
where \(P_{\rm EP}\) is expressed in Pascal units.
In Figure \ref{fig:polytrope}, we show this behavior for the different portions of the SEP event marked in Figure \ref{fig:overview}.
These portions are marked as pre-ICME (solar wind upstream of CME-driven shock), sheath (downstream of shock), CME ejecta (magnetic obstacle), and post-ICME (solar wind after ICME).
Each portion of the SEP event appears to have different polytropic behaviors with respect to each other.
The polytropic indices and their standard error for each portion are given in Table \ref{tab:thermo_params}, which are consistent with sub-isothermal polytropic processes, \(\gamma_{\rm EP} < 1\), supporting the observations of anti-correlated derived temperature and density.
While the overall trends in Figure \ref{fig:polytrope} are linear, we note that there are breaks from linearity in the sheath and CME ejecta regions, possibly the result of encountering separate magnetic structures, warranting further investigation.

\newpage

{
\begin{table}[!ht]
    \makebox[.9\textwidth]{%
    \begin{tabular}{|c|c|c|c|c|}
        \hline
        Parameter & pre-ICME & Sheath & CME Ejecta & post-ICME \\
        \hline
        Polytropic Index, \(\gamma_{\rm EP}\) & 0.32 \(\pm\) 0.01 & 0.47 \(\pm\) 0.02 & 0.75 \(\pm\) 0.01 & 0.90 \(\pm\) 0.01 \\
        \hline
        Entropic Coefficient, \(\eta_{\rm EP}\) & 9.24 \(\pm\) 1.03 & -3.71 \(\pm\) 0.26 & 1.18 \(\pm\) 0.27 & 1.74 \(\pm\) 0.39 \\
        \hline
    \end{tabular}
    }
    \caption{ODR fitted values and standard deviations of the EP polytropic index (\(\gamma_{\rm EP}\)) and entropic coefficient (\(\eta_{\rm EP}\)) for each portion of the 15 Feb 2022 SEP event.}
    \label{tab:thermo_params}
\end{table}
}

\begin{figure}[ht]
    \centering
    \includegraphics[width=.6\textwidth]{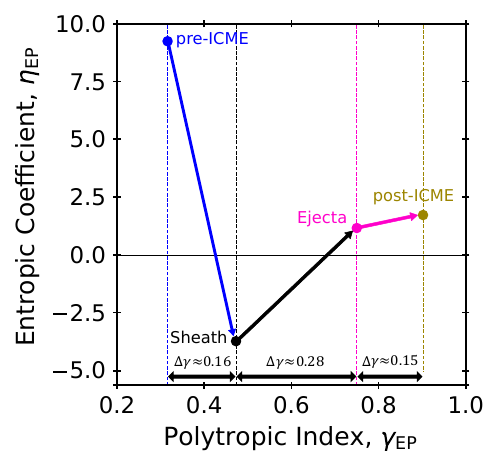}
    \caption{The evolution of the EP entropic coefficient (\(\eta_{\rm EP}\)) with respect to the EP polytropic index (\(\gamma_{\rm EP}\)) between each portion of the 15 Feb 2022 SEP event.}
    \label{fig:eta_gammap}
\end{figure}

In Figure \ref{fig:overview}, the temperature and kappa are generally increasing and positively correlated, such that the entropy of the system is also increasing. 
For this, in Figure \ref{fig:entropy}, we show an estimate of the entropic measure \(S_{\rm EP}\), a function of temperature and density, that is statistically correlated with \(\kappa_{\rm EP}\) given by an ODR linear fit of the form \(S_{\rm EP} \propto \eta_{\rm EP} \cdot \kappa_{\rm EP}\), where \(\eta_{\rm EP}\) is the entropic coefficient (values of \(\eta_{\rm EP}\) are given in Table \ref{tab:thermo_params}).
We report that in the regions pre-ICME and post-ICME, the entropic measure is increasing monotonically with \(\kappa_{\rm EP}\). 
The supports the idea that a kappa distribution with a steeper tail (larger spectral index or larger \(\kappa_{\rm EP}\)) corresponds to a wider thermal distribution (larger average temperature), leading to an increased entropic measure. 
However, in the regions of the sheath and CME ejecta, the quantities \(S_{\rm EP}\) and \(\kappa_{\rm EP}\) are anti-correlated and uncorrelated, respectively. 
This suggests that the presence of the shock and CME tends to reduce or conserve the entropic measure of energetic particles even though the value of thermodynamic \(\kappa_{\rm EP}\) continues to increase throughout these regions.  
Values of \(\gamma_{\rm EP}\) and \(\eta_{\rm EP}\) for each portion of the SEP event are compiled in Figure \ref{fig:eta_gammap}, revealing a dependence in the trend of entropy with respect to the polytropic index.
Furthermore, the heating processes of a particle population can be indicated by the variation of the polytropic index among different plasma structures.
These variations can mark transitions in the undergoing thermodynamic behavior of the particles.
In the case of the ICME examined in this study, one can investigate the polytropic displacement (\(\Delta \gamma_{\rm EP}\)) between each adjacent portion of the ICME, which can be connected to the gradient of turbulent energy, \(\epsilon_t\), over radial distance \(R\) as follows:

\begin{equation}
    \Delta \gamma_{\rm EP} \sim \frac{\Delta \epsilon_t / \langle E\rangle}{\Delta R/R}
\end{equation}
where \(\langle E \rangle\) is the thermal energy \citep{DayehLivadiotis2022ApJL_PolytropicBulk_ICMEs}.
Although the radial distance covered by PSP during this SEP event is small, the present results suggest that the polytropic index is monotonically increasing such that the net \(\Delta \gamma_{\rm EP}\) is positive.
This motivates a systematic analysis to reveal the behavior of the gradients of \(\gamma_{\rm EP}\) and \(\epsilon_t\) as a function of radial distance.


\section{Conclusion} \label{sec:conclusion}

In this paper, we derived the thermodynamic parameters of EP density, temperature, and kappa, characterizing 10 to 60~MeV solar energetic protons over the entirety of a well-sampled SEP event, for the first time, using PSP/IS\(\odot\)IS EPI-Hi HET-A proton observations.
We found that the tail of an SEP population is well-described by a kappa distribution function. 
Particularly, in this high energy limit (\(\epsilon \gg \kappa_0 T_{\rm EP}\)), a kappa distribution reduces to a power-law tail, whose fitting parameters involve the thermodynamic parameters of density, temperature, and kappa. 
This functional dependence of the power-law tail fitting parameters on the thermodynamic parameters comprises a special signature expression, characteristic to the formulation of the kappa distribution.
This signature expression can be used for examining whether the tail originated from a kappa distribution function, and if so, to proceed to determine the thermodynamic parameters of this distribution. 
This technique was previously developed by \citet{LivadiotisMcComas2011ApJ_InvariantKappa,LivadiotisMcComas2012ApJ_KappaThermoProcesses,LivadiotisMcComas2013SSRv_KappaDist_SpSc,LivadiotisMcComas2023ApJ_PolytropicIndex_Heating} and later adapted by \citet{LivadiotisEA2024_SEPthermo_technique} to SEPs.

We showed the temporal evolution of derived EP temperature, density, thermal pressure, and spectral index, that is, equivalent to the thermodynamic parameter kappa, \(\kappa_{\rm EP}\) (Figure \ref{fig:overview}), throughout each portion of the SEP event (pre-ICME, sheath, CME ejecta, and post-ICME). 
We also provided the first insights of the polytropic and entropic behavior of 10 to 60~MeV solar energetic protons.

Overall, we found the following thermodynamic properties of solar energetic protons:
\begin{enumerate}
    \item The thermodynamic parameter \(\kappa_{\rm EP}\) varies smoothly throughout each region of the SEP event. It increases into the CME-driven shock and CME ejecta where it reaches its peak value. Afterwards, it is found to gradually decrease throughout the post-ICME region.
    \item The EP temperature and density are highly variable in the pre-ICME, sheath, and beginning half of the CME ejecta regions, after which they become more smoothly varying into the post-ICME region. Nonetheless, their variability is anti-correlated, consistent with sub-isothermal polytropic processes (polytropic index lower than one).
    \item The EP temperature and kappa values are usually positively correlated, leading to an increasing entropic measure. This holds true in the pre-ICME, CME ejecta, and post-ICME regions. However, the EP entropic coefficient is negative for the sheath region, suggesting a reduction of entropy with increasing \(\kappa_{\rm EP}\).
\end{enumerate} 
Finally, we conclude that SEPs are characterized by a complex and evolving thermodynamic behavior, consisting of multiple sub-isothermal polytropic processes and a large-scale trend of increasing temperature, kappa, and entropy.

The procedure demonstrated here enables the space physics community to investigate the thermodynamic properties of any particle population where measurements of EPs are available, agnostic to specific missions.
Future analyses with the intention of automatic detection of magnetic flux tubes may use these procedures to assist in determining intervals of similar magnetic connectivity to EP acceleration regions during solar events. 
EP thermal pressures of SEPs may provide insight into the thermal energy, which can be used to determine the contribution of SEPs to the energy budget of observed solar events, such as solar flares and CMEs. These procedures could potentially help determine whether different acceleration mechanisms can be detected by the differences in the accelerated SEP thermodynamic properties.
Additionally, improvements to the technique to include relativistic effects may be of interest for the application of this technique to even higher energy particles.
Thus, this study and its companion study by \citet{LivadiotisEA2024_SEPthermo_technique}, open a new area and set of procedures for analyzing SEP observations in the context of their thermodynamic behavior and collective processes. \newline \newline


We thank thank the IS\(\odot\)IS team and everyone that made the PSP mission possible. The IS\(\odot\)IS data and visualization tools are available to the community at \href{https://spacephysics.princeton.edu/missions-instruments/PSP}{https://spacephysics.princeton.edu/missions-instruments/PSP}. 
PSP was designed, built, and is operated by the Johns
Hopkins Applied Physics Laboratory as part of NASA’s Living with a Star (LWS) program (contract NNN06AA01C).
This research was partially funded by PSP GI grant 80NSSC21K1767.


%



\newpage
\appendix
\counterwithin{figure}{section}
\section{Entropic Measure}\label{app:entropy}

\begin{figure}[!ht]
    \centering
    \includegraphics[width=.6\textwidth]{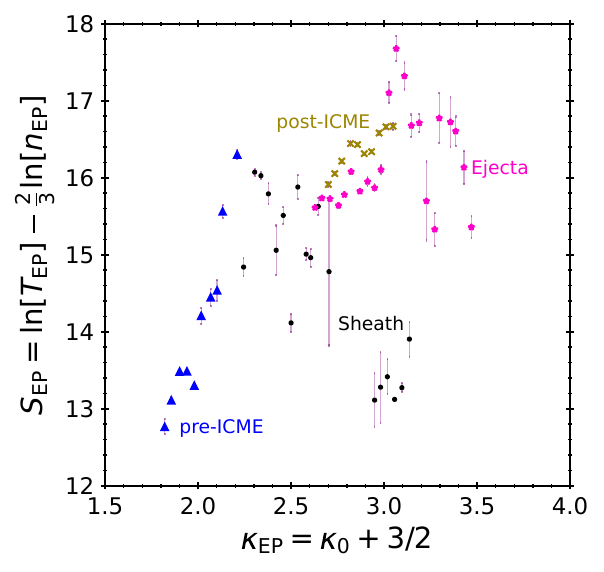}
    \caption{The entropic measure \(S_{\rm EP}\) as a function of thermodynamic \(\kappa_{\rm EP}\) (error bars representing standard error) for each portion of the 15 Feb 2022 SEP event marked along the top axis in Figure \ref{fig:overview}. These points represent the weigthed average of equidistant bins of size 0.04 along \(\kappa_{\rm EP}\).}
    \label{fig:entropy}
\end{figure}

In Figure \ref{fig:entropy} we provide the evolution of the EP entropic measure, \(S_{\rm EP}= \ln T_{\rm EP} - \frac{2}{3} \ln n_{\rm EP}\) as a function of the thermodynamic \(\kappa_{\rm EP}\), for each portion of the 15 Feb 2022 SEP event. 
For the regions pre-ICME and post-ICME, the entropic measure increases, given by a positive value for the entropic coefficient, \(\eta_{\rm EP}\) (see Figure \ref{fig:eta_gammap} and Table \ref{tab:thermo_params}). 
This is expected, since an increase in \(\kappa_{\rm EP}\), or a steeper spectral index, leads to a wider thermal distribution reflecting a state of higher entropy, or closer to thermal equilibrium.
However, in the regions of the sheath and CME ejecta, the entropy coefficient is negative and nearly zero, suggesting a reduction of entropy or its conservation, respectively.

\newpage





\end{document}